\documentclass[sts]{imsart}

\startlocaldefs
\usepackage{amssymb, latexsym, amsthm, amsmath, lineno, epsfig, mathtools, hyperref, todonotes, booktabs}
\newcommand\given{{\,|\,}}
\endlocaldefs

\begin{document}

\begin{frontmatter}

\title{J.B.S.\ Haldane Could Have Done Better}
\runtitle{J.B.S.\ Haldane Could Have Done Better}

\thankstext{t1}{CV was supported by the Austrian Science Fund (FWF): DK W1225-B20.} 

\author{\fnms{Claus} \snm{Vogl}\ead[label=e1]{claus.vogl@vetmeduni.ac.at}}
\address{\printead{e1}}
\affiliation{Veterin\"armedizinische Universit\"at Wien}
\address{ Institut f\"ur Tierzucht und Genetik, Veterin\"armedizinische Universit\"at Wien, Veterin\"arplatz 1, A-1210 Vienna, Austria}

\runauthor{C.\ Vogl}




\end{frontmatter}

\section*{Comment on: ``J.B.S.\ Haldane's contribution to the Bayes factor hypothesis test'' by Etz and Wagenmakers}

Etz and Wagenmakers \cite{Etz17} (and an earlier version of this paper available at: https://arxiv.org/abs/1511.08180) review the contribution of J.B.S. Haldane to the development of the Bayes factor hypothesis test. They focus particularly on Haldane's proposition of a mixture prior in his first example on genetic linkage mapping in the Chinese primrose {\em (Primula sinensis)\/} \cite{Hald32}. As Haldane never followed up on these ideas, it is difficult to gauge his motivation and intentions. Haldane himself states his purpose in the beginning of the article \cite{Hald32}: 

\begin{quote} Bayes theorem is based on the assumption that all values in the neighborhood of that observed are equally probable {\it a priori}. It is the purpose of this article to examine what more reasonable assumptions could be made, and how it will affect the estimate given the data.
\end{quote}

Compactly restated: {\em flat priors should be replaced by more reasonable assumptions.} But I will argue that in the very same article, in the very first example, Haldane himself uses a flat prior instead of a more reasonable prior. 

\paragraph{Haldane's primrose example with a flat prior.} The data come from a (hypothetical) observation of 400 meioses in the primrose; 160 of them are cross-overs. Let $\rho$ be the recombination rate between the two loci. The likelihood is a binomial
\begin{equation}
    p(y=160\given\rho,N=400)=\binom{400}{160}\,\rho^{160}(1-\rho)^{240}\,.
\end{equation}
Haldane argues that {\it P.~sinensis} has twelve chromosomes of about equal length. Recombination between unlinked loci on different chromosomes is free, such that the recombination rate $\rho=\frac12$. This is reflected in Haldane's prior by a point mass of $\frac{11}{12}$ on $\rho=\frac12$. With probability $\frac{1}{12}$, the two loci reside on the same chromosome, i.e., the two loci are linked. Conditional on linkage, Haldane assumes $0\leq\rho<\frac12$ and a {\em flat prior} of $p(\rho)=2$, such that his marginal posterior distribution becomes
\begin{equation}\label{eq:exact}
 p(y=160\given N=400)=\frac1{6} \binom{400}{160}
 \int_0^{\frac{1}{2}}\rho^{160}(1-\rho)^{240}\,dx\,.
\end{equation}
He continues to approximate by extending the upper integration limit to one
\begin{equation}
\begin{split}\label{eq:approx}
p(y=160\given N=400)&\approx\frac1{6} \binom{400}{160}\int_0^{1} \rho^{160}(1-\rho)^{240}\,dx\\
&=\frac1{6} \binom{400}{160}\frac{160!\,240!}{401!}=\frac{1}{6\cdot401}\,.   
\end{split}
 \end{equation}
But the {\em flat prior is unreasonable}, given Haldane's knowledge of genetic linkage. 

\begin{figure}[ht]
\includegraphics[width = 12cm]{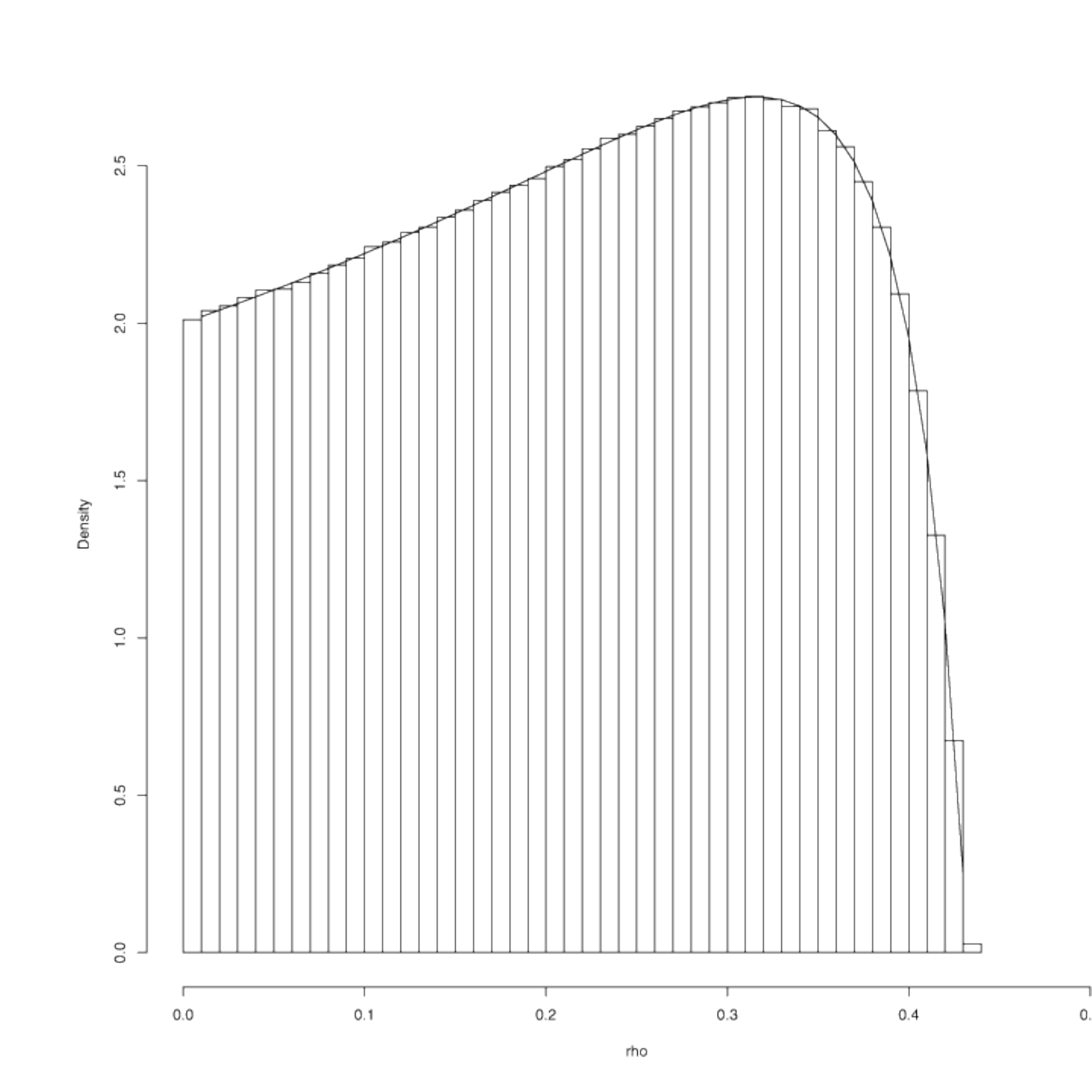}
\caption{The prior distribution of $\rho$ given $L=1$ and assuming equal distribution of positions and Haldane's mapping function. The histogram is produced from a simulation; the solid line corresponds to the distribution in eq.~(\ref{eq:prior_rho}).}\label{fig:prior_rho}
\end{figure}

\paragraph{A better prior.} Chromosomes are one-dimensional structures on which loci reside. The recombination rate $\rho$ between two genes is a function of their distance on the chromosome. It would have been reasonable for Haldane to assume that a locus can be located anywhere on a chromosome with equal probability and that the locations of two loci are independent of each other. Then the genetic distance $x$ in units of proportions of the length of the chromosome (denoted with $L$ and measured in cross-over rates, i.e.,  Morgan) between the two loci would be given by a beta 
\begin{equation}
 p(x)=\frac{\Gamma(3)}{\Gamma(1)\Gamma(2)}\,x^{1-1}(1-x)^{2-1}\,.
\end{equation}
Haldane \cite{Hald19} himself derived a bijective function that maps genetic distance $x$ to recombination rate $\rho$: 
\begin{equation}
    \rho_{\given x,L}= \frac{1-e^{-2Lx}}{2}\,.
\end{equation}
The number of cross-overs per meiosis per chromosome is about one, a fact probably known to Haldane, such that I set $L=1$. Changing variables from $x$ to $\rho$, the prior distribution of $\rho$ then becomes
\begin{equation}\label{eq:prior_rho}
    p(\rho)=\frac{2+\log(1-2\rho)}{1-2\rho}
\end{equation}
with $0\leq\rho\leq \frac{1-e^{-2}}{2}$ (Fig.~\ref{fig:prior_rho}). Note that, for the primrose example, the maximum likelihood estimator of the recombination rate is $\hat \rho=160/400=0.4$. In this parameter region the prior (\ref{eq:prior_rho}) differs considerably from the flat prior $p(\rho)=2$.  

\paragraph{Speculations on Haldane's intentions.} Haldane most certainly also went through the above considerations; after all, he himself developed a very useful mapping function. Reading the article carefully, I consider its main purpose not the mixture prior in the primrose example, but rather the investigation of different parameter regions of the binomial and its conjugate distribution, the beta. The primrose example is in a parameter region, where probabilities of failure and success are about equal. (Realize that the example data are actually closer to equal probabilities than is usually encountered in linkage studies, where sample sizes are often about 50 to 100, rather than Haldane's 400, which would have made detection of linkage unlikely with a true $\rho=0.4$.) For this, a flat prior is reasonable, i.e., a prior beta with $\alpha=\beta=1$. Haldane may actually have been more interested in the approximate distribution~(\ref{eq:approx}) than in the exact one~(\ref{eq:exact}). The other examples in Haldane's article pertain to parameter regions where success (or failure) probabilities are close to zero or one. Then a flat prior would put too much weight into the middle of the parameter region and a prior with $\alpha\to0$ and $\beta=1$ proportional to $\frac{1}{\rho}$, or with $\alpha=\beta\to0$ proportional to $\frac{1}{\rho(1-\rho)}$, would be preferable. In this light, a more complicated prior distribution than the beta and its asymptotes would have been useless, even though Haldane could have derived it easily. I thus believe that, for the sake of generality, Haldane chose to not do better than flat in the primrose example. Furthermore, I agree with Etz and Wagenmakers \cite{Etz17}: 

\begin{quote}It was the specific nature of the linkage problem in genetics that caused Haldane to serendipitously adopt a mixture prior comprising a point mass and smooth distribution.\end{quote} 

\paragraph{A genetic red herring.} Modern genetics has shown cross-over rates to be variable along a chromosome, with low rates at the chromosome ends and around the centromere. Since the distribution of genes on chromosomes also follows roughly the same pattern, this complication can be ignored, as long as genetic position is based on mapping distances (in units of Morgan) and not physical distances (in units of basepairs).

\section{Acknowledgments}
I thank Alexander Etz and Eric-Jan Wagenmakers for inspiration and encouragement. My research was supported by the Austrian Science Fund (FWF): DK W1225-B20.

\bibliography{bibliography}
\bibliographystyle{imsart-number}
\newpage

\section*{Figures}

\end{document}